\documentclass{aa}
\usepackage[dvips]{graphics}
\begin{document}

   \thesaurus{01 (11.14.1; 11.19.1)}
   \title{The effect of bars on the obscuration of active nuclei}

   \subtitle{}

   \author{R. Maiolino \inst{1} \and
			G. Risaliti \inst{2} \and
          M. Salvati \inst{1}
          }

   \offprints{R. Maiolino}

   \institute{Osservatorio Astrofisico di Arcetri, Largo E.Fermi 5,
              I-50125, Firenze, Italy, maiolino@arcetri.astro.it,
              salvati@arcetri.astro.it
	\and
		Dipartimento di Astronomia, Universit\`a di Firenze, Largo E.Fermi 5,
			I-50125, Firenze, Italy, risaliti@arcetri.astro.it
             }

   \date{Received / Accepted }

   \maketitle

   \begin{abstract}
We present evidence for a strong correlation between the gaseous
absorbing column
density towards type 2 Seyfert nuclei
and the presence of a stellar bar in their host galaxies.
Strongly barred Seyfert 2 galaxies have an average N$_H$ that is two orders
of magnitude higher than non-barred Sy2s. More than 80\% of Compton thick
(N$_H>10^{24}cm^{-2}$) Seyfert 2s are barred. This result indicates that
stellar bars are effective in driving gas in the vicinity of
active nuclei.
   \end{abstract}
%

\section{Introduction} \label{intro}

Large scale gravitational
torques, such as bars and interactions, are thought to transport gas into
the central region of galaxies and, specifically, in the vicinity
of active galactic nuclei (Shlosman et al. 1990).
Seyfert galaxies are the low luminosity subset of AGNs. According
to the accreting supermassive black hole paradigm, the accretion
rate inferred for Seyfert nuclei is low ($10^{-1}-10^{-2}M_{\odot}
yr^{-1}$) and, therefore, much fuelling from the host galaxy is not
required. Indeed, McLeod \& Rieke (1995),
Ho et al. (1997a), Mulchaey \& Regan (1997) show that the
occurrence of bars in Seyfert galaxies is not higher than in 
normal galaxies.
Yet, there is both theoretical 
and observational evidence that stellar bars do drive gas into the central
region of galaxies (Athanassoula 1992, Tacconi et al. 1997, Laine et al.
1998). The resulting central concentration of gas might not be relevant for
the fuelling process, but can play a role in the obscuration of the active
nucleus, i.e. stellar bars could contribute to the obscuration
that affects Seyfert 2s. This connection would be very important for the
unified theories (Antonucci 1993).
In this letter we tackle this issue by comparing the
degree of obscuration in Sy2s with the strength of the stellar bar in their
host galaxy.

\section{Dependence of the absorbing N$_H$ on the bar strength} \label{nh_bar}

Hard X-ray spectra can be regarded as the best tool to measure directly
the absorbing column density in Seyfert galaxies. Recent surveys have
significantly enlarged the sample of Sy2s for which an estimate of N$_H$ is
available, and have also reduced the bias against heavily obscured objects
that plagued former studies (Maiolino et al. 1998a, Bassani et al. 1998,
Risaliti et al. 1998).

We restricted our study to obscured Seyferts in the Maiolino \& Rieke (1995)
sample, completed with 18 additional objects discovered in the Ho et al.
(1997b) survey. These two Seyfert samples are much less
biased than others in terms of luminosity and obscuration of the active
nucleus and, therefore, can be considered representative of the local
population of Seyfert galaxies.

For AGNs which are thin to Compton scattering
(i.e. N$_H < 10^{24}cm^{-2}$) along our line of sight the N$_H$ can be
derived from the photoelectric cutoff in the 2--10 keV spectral range,
provided that the signal to noise is high enough. If the source is Compton
thick then the direct component in the 2--10 keV range is completely
suppressed and we can only observe the reflected component, generally
little absorbed. As a consequence, Compton thick Sy2s are sometimes
misidentified as ``low absorption objects'' when observed in the 2-10 keV
range. However, the fact that the absorbing column density is actually
higher ($>10^{24}cm^{-2}$) can be inferred from several indicators, such as
the equivalent width of the Fe line and the spectral slope, and by comparing
the X-ray flux with other isotropic indicators of the intrinsic luminosity.
A more detailed discussion of this issue is given in
Maiolino et al. (1998a) and Bassani et al. (1998).
We just mention that sensitive spectra at higher energies, such as
those obtained by BeppoSAX in the 10--200 keV band,
can identify column densities in the range $10^{24}-10^{25}cm^{-2}$
(Matt et al. 1998, Cappi et al. 1998) or set a lower
limit of $10^{25}cm^{-2}$ (Maiolino et al. 1998a).

\begin{figure}[!t]
\resizebox{8truecm}{!}{\includegraphics{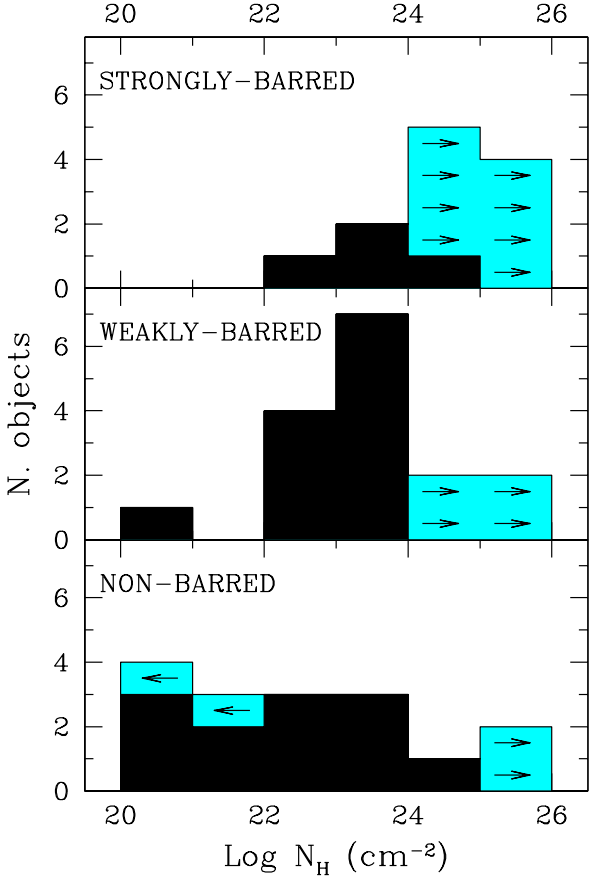}} 
 \caption{Distribution of the absorbing column density in Sy2s as a
function of the bar strength}
\label{fig_barnh}
\end{figure}

\begin{table}
\caption[]{Absorbing N$_H$ and bar classification for the
Sy2s sample.}\label{tab_obs}
{\tiny
\begin{center}
\begin{tabular}{lcccc}
\hline
\hline
Name & Log(N$_H$)$^a$ & \multicolumn{3}{c}{Bar classification$^b$} \\
     &         & RC3 & Others$^c$ & Adopted \\
\hline
NGC1068 & $>$25 & SA & bar$^1$ & SB \\
NGC1365 & 23.3  & SB & -- & SB\\
NGC1386 & $>$24 & SB & -- & SB \\
NGC1808 & 22.5  & SAB & -- & SAB\\
NGC2110 & 22.5  & SAB & -- & SAB \\
NGC2273 & $>$25 & SB & double-bar$^2$ & SB\\
NGC2639 & 23.6  & SA & no-bar$^3$ & SA\\
NGC2992 & 21.8  & -- & no-bar$^5$ & SA \\
NGC3031 & 20.9 & SA & -- & SA\\
NGC3079 & 22.2 & SB  & -- & SB\\
NGC3081 & 23.8  & SAB & double-bar$^{2,7}$ & SAB\\
NGC3147 & 20.6  & SA & -- & SA\\
NGC3281 & 23.9  & SAB & -- & SAB\\
NGC3393 & $>$25 & -- & bar$^{2,5}$ & SAB\\
NGC4258 & 23.2  & SAB & -- & SAB\\
NGC4388 & 23.6 & SA & boxy-bulge$^{4,5}$ & SAB \\
NGC4507 & 23.5  & SAB & bar$^2$ & SAB\\
NGC4565 & $<$21.8  & SA  & -- & SA\\
NGC4579 & 20.6  & SAB & -- & SAB\\
NGC4594 & 21.7  & SA & -- & SA\\
NGC4941 & 23.6  & SB & -- & SB\\
NGC4939 & $>$25 & SA & -- & SA \\
NGC4945 & 24.6  & SB & -- & SB\\
NGC5005 & $>$24 & SAB & -- & SAB\\
NGC5033 & 20.9 & SA & no-bar$^3$ & SA\\
NGC5135 & $>$24 & SB & bar$^{2,3}$ & SB \\ 
NGC5194 & 23.7  & SA & -- & SA\\
NGC5347 & $>$24 & SB & bar$^2$ & SB\\
NGC5506 & 22.5  & -- & no-bar$^3$ & SA\\
NGC5643 & 23.3  & SAB & bar$^2$ & SAB\\
NGC5674 & 22.8  & SAB & bar$^4$ & SAB\\
NGC7172 & 22.9  & --  & no-bar$^3$ & SA\\
NGC7314 & 22.1  & SAB & bar$^3$ & SAB \\
NGC7319 & 23.5  & SAB & -- & SAB\\
NGC7582 & 23.1  & SB & -- & SB\\
NGC7590 & $<$20.9 & -- & no-bar$^3$ & SA\\
IC2560  & $>$24 & SB & -- & SB\\
IC3639  & $>$25 & SB & bar$^2$ & SB\\
IC5135  & $>$24 & -- & bar$^{2,3}$ & SAB\\
IC5063  & 23.4  & SA & no-bar$^2$ & SA\\
Circinus & 24.7 & SA & -- & SA \\
IRAS0714 & $>$25 & SA & -- & SA \\
IRAS1832-59 & 22.1 & SA & -- & SA \\
Mk1066  & $>$24 & SB & -- & SB\\
\hline
\end{tabular}
\end{center}
\begin{list}{}{}
\item $^a$ In units of cm$^{-2}$; these values are from
  Bassani et al. (1998) and Risaliti et al. (1998).
\item $^b$ SA = non-barred; SAB = weakly barred; SB = strongly barred.
\item $^c$ References: 1--Thatte et al. (1997); 2--Mulchaey et al. (1997);
3--Hunt et al. (in prep.);
4--McLeod \& Rieke (1995); 5--Alonso-Herrero et al. (1998).
\end{list}
}
\end{table}

In Tab. 1 we list all the Seyfert 2, 1.9 and 1.8 galaxies (i.e. those showing
indication of obscuration) in the joint Maiolino \& Rieke (1995) and Ho et al.
(1997b) samples which have an estimate of the
absorbing N$_H$ based on X-ray data, and for which a bar
classification is also available. With regard to the
bar we generally favored the classification based on near--IR
images (since these are less affected by extinction) and quantitative
identifications of the stellar bars based on ellipticity, position
angle, ``boxy-shape'' and brightness profile arguments
as measured from digital (unsaturated) images.
Otherwise the optical classification reported in the RC3 was adopted
(de Vaucouleurs et al. 1991).
However, in almost all cases the RC3 identification of bars was in agreement
with the other works.
Whenever the RC3 classification was not in conflict with the near-IR and/or
``quantitative'' classification (i.e. in most cases), we further split
barred systems in strongly barred (SB) and weakly barred (SAB) as reported in
the RC3. Four objects not classified or classified as non-barred
in the RC3, while their infrared images show indications of a bar; in these
cases we adopt the classification SAB, with the exception of NGC1068, whose
bar appears strong in the K-band image (Thatte et al. 1998).

The completeness of this Sy2s sample is discussed in detail in
Risaliti et al. (1998). We could not find obvious biases or other
correlations that could introduce spurious relations between N$_H$ and
bar properties. In particular, there is no correlation
between N$_H$ and the luminosity of active nuclei (Risaliti et al. 1998).

Fig. 1 shows the distribution of N$_H$ for subsamples of increasing stellar
bar strength, ordered from bottom to top.
There is a clear tendency for
N$_H$ to increase along the sequence. Tab. 2 quantifies this apparent
trend. The median of the N$_H$ distribution\footnote{The median is
estimated by means of the Kaplan-Meier estimator that takes into account also
censored data.} increases by more than two orders of magnitude going from
unbarred to strongly barred Sy2s. The confidence of the results is given
by the Gehan test (Feigelson \& Nelson 1985),
that takes into account also censored data:
strongly barred and unbarred Sy2s have N$_H$ distributions
 that are different at a confidence
level higher than 99\%. Another impressive result is that more than
80\% of Compton thick
(N$_H > 10^{24}cm^{-2}$) Sy2s are barred (13 out of 16), to be compared with
$\sim$ 55\% of the general population and specifically of early type systems
that typically host Seyfert activity (Sellwood \& Wilkinson 1993,
Ho et al. 1997a). Also 56\% of Compton thick Sy2s are hosted in
strongly barred systems, to be compared with $\sim$ 25\% of the general
population.

\begin{table}[!t]
\caption[]{Properties of the N$_H$--bar correlation in Sy2s}
\label{tab_res}
\begin{tabular}{lccc}
\hline
\hline
Parameter&Non--barred&Weakly &Strongly \\
 & &barred&barred\\
\hline
No. objects& 16 & 16 & 12 \\
Med. Log(N$_H$)$^a$& 22.1 & 23.5 & 24.4 \\
Compton thick:& & & \\
\hskip0.5truecm No. (\%)& 3 (18.7\%) & 4 (25.0\%) & 9 (56.2\%) \\
Probability$^b$ &  & 89\%& $>$99\% \\
\hline
\end{tabular}
 $^a$~Median of Log(N$_H$) in units of cm$^{-2}$.\\
 $^b$~Probability for the N$_H$ distribution of intermediate and
strongly barred Sy2s to be different from the N$_H$ distribution
of non-barred Sy2s.
\end{table}

\section{Other indications}\label{other}

Maiolino et al. (1997) found that Sy2s are characterized by a rate of
non-axisymmetric potentials (including interactions and peculiar
morphologies) about 20\% higher than Sy1s, this difference appears
significant.
Hunt \& Malkan (1998) find the
occurence of bars in Sy2s not significantly higher than in Sy1s within the
CfA and the 12$\mu m$ samples. In
the samples of Ho et al. (1997a) and Mulchaey \& Regan (1997) the occurrence
of bars in type 2 Seyferts is 10--20\% higher than type 1 Seyferts. These
results indicate that even if bars drive gas into the circumnuclear region,
such gas does not reduce much the opening angle of the light cones.
Yet, if the correlation between bar strength and amount of
circumnuclear gas obtained for Sy2s applies also to Sy1s, we would expect 
a large amount of gas in the circumnuclear region of barred Sy1s as well.

The circumnuclear (cold) gas is
expected to Compton-reflect the nuclear X-ray radiation. This Compton-reflected
component should flatten the X-ray spectrum in the 10--30 keV spectral range.
Therefore, within the bar--circumnuclear gas connection scenario depicted
above, we would expect barred Sy1s to have a flatter spectrum in the
10--30 keV band. However, this test is subject to various caveats. First,
variability affects the slope of the observed spectrum because of the
time-lag between the primary and reprocessed radiation.
Second, a fraction
of the ``cold'' reflection is expected to come from the accretion disk.
Third, the effect is expected to be small: the
reflected component should contribute no more than $\sim$30\% in this spectral
region. Fourth, to date spectra in this X-ray band are sparse and with low
sensitivity.
So far, the only (small) sample of Sy1s observed at energies
higher than 10 keV is the one presented in Nandra \& Pounds (1994), that use
Ginga data. Their sample contains nine Sy1s whose host galaxy have a
bar classification. As shown in Tab. 3 the spread of the photon index
measured between 10 and 18 keV is large. Nonetheless,
Tab. 3 shows a tendency for the
hard X-ray spectra of barred Sy1s to be flatter than the unbarred Sy1s.

\begin{table}[!b]
\caption[]{Properties of Sy1s as a function of the bar strength}
\label{tab_sy1}
\begin{tabular}{lccc}
\hline
\hline
Parameter&Non--barred&Weakly &Strongly \\
 & & barred & barred\\
\hline
$\langle
\Gamma _{10-18 keV}\rangle$& 1.69$\pm$ 0.24 & 1.31$\pm$0.10 & 1.24$\pm$0.47 \\
(No. objects) & (3) & (3) & (3) \\
\hline
$\langle
log(L_{IR}/L_X)\rangle ^a$&
0.92$\pm$ 0.07 & 1.20$\pm$0.23 & 1.58$\pm$0.80 \\
(No. objects) & (5) & (6) & (8) \\
\hline
\end{tabular}
$^a$~$L_{IR}$ = N band ($\sim 10\mu m$) luminosity; $L_X$ = 2--10 keV
 luminosity.
\end{table}

Large amounts of circumnuclear gas in Sy1s could be detected via
dust-reprocessed light in the infrared. More circumnuclear gas would imply
more warm (AGN-heated) dust, hence more infrared emission relative to the
intrinsic luminosity of the AGN (traced by the hard X-ray luminosity).
The mid--IR
($\sim 10\mu m$) is an excellent band to look for this excess, since
the AGN IR emission peaks there (Maiolino et al. 1995).
Also, by using narrow beam photometry
it is possible to isolate the contribution of the AGN from the host galaxy.
Within the scenario of the bar--circumnuclear gas connection,
barred Sy1s are expected to show a mid-IR to X-ray flux ratio higher than
non-barred Sy1s. Yet, several caveats affect this test as well. Both short
and long term variability plague the reliability of the X-ray flux as a
calibrator of the AGN luminosity.
Equilibrium temperature arguments indicate that
the dust emitting significantly at 10$\mu m$ should be located within the
central 1--10 pc; therefore, excess of circumnuclear gas
distributed over the 100 pc scale would not be probed by this indicator.
Finally, even the small aperture (5$''$) used in most of the groundbased
mid-IR observations might include the contribution from a central compact
starburst. Tab. 3 reports the mean of the 10$\mu m$/2--10keV luminosity ratio
as a function of the bar strength for a sample of 19 Sy1s. The
10$\mu m$ data are from Maiolino et al. (1995)
and from Giuricin et al. (1995); the
X-ray data are from Malaguti et al. (1994),
where we choose the datum closest in time to the mid-IR observation, to
minimize long term variability effects. There is a tendency for barred
systems to have a higher L$_{10\mu m}$/L$_X$ ratio, though the spread is large
and the statistics are poor.

\section{Discussion}\label{disc}
 
The important result of our study is that the absorbing column density
of type 2 Seyferts strongly correlates with the presence of a stellar
bar in their host galaxies.
As discussed in the Introduction, this result is not completely
unexpected: stellar bars are very effective in driving gas into the central
region, thus contributing to the obscuration of AGNs. On the other hand this
gas should not play a major role in powering low luminosity AGNs, such as
Seyfert nuclei, given the lack of correlation between stellar bars and
Seyfert activity (McLeod \& Rieke 1995, Mulchaey \& Regan 1997, Ho et al.
1997a).

As discussed in the former section,
the fraction of bars
in Sy2s is only moderately higher (at most by 20\%) than in Sy1s.
So, the gas accumulated
in the central region by the stellar bar increases the column density outside
the light cones, but it does not increase much the covering factor
of the obscuring material. A possible explanation is that
radiation and wind/jet pressure
act to destroy or expel molecular clouds that happen to enter the light
cones, while outside the light cones self-shading allows molecular clouds to
survive and
pile up along our line of sight. Another possibility is that the inflowing gas
concentrates at dynamical resonances (eg. Lindblad resonances)
forming obscuring
tori. The solid angle that such tori subtend to the AGN depends on their
inner radius and thickness, that in turn depend on the dynamical and 
kinematical properties of the nuclear region, but are relatively independent
 of the amount
of gas in the torus (if the gas self--gravity is significant the
torus would flatten, thus actually reducing its covering solid angle).
On the other hand, the N$_H$ through the torus depends linearly on its
mass.
 
Our result has also important implications on the scales over which the
obscuring material is distributed. Large scale stellar bars can transport
gas into the central few 100 pc, but they exert little influence on the
gas dynamics on smaller nuclear scales. As a consequence, a first
interpretation of our bar--N$_H$ link is that a large fraction of the
obscuration in Sy2s occurs on the 100 pc scale. This is in line with results
obtained from HST images, that ascribe the obscuration of several
Sy2s nuclei to dust lanes or disks a few 100 pc in size (eg.
Malkan et al. 1998).
Yet, one of the most interesting results of our study is that
stellar bars appear very effective in making the obscuration of Sy2 nuclei
so high to be Compton thick: more than 80\% of Compton thick Sy2s are
barred. As discussed in Risaliti et al. (1998),
column densities larger than 10$^{24}cm^{-2}$ are unlikely to be distributed
on the 100 pc scale, since the implied nuclear gas mass would exceed
the total dynamical mass in the same region for several objects.
This consideration generally constrains most of the Compton thick gas to be
located within a few 10 pc from the nucleus. On the other hand,
on these small scales the gas dynamics is expected to be little affected by
the non-axisymmetric potential of a stellar bar in the host galaxy.
The connection between large
stellar bars and Compton thickness of Sy2s thus
requires some mechanism to link the dynamics on these different scales.
Shlosman et al. (1989) proposed that if the large scale
stellar bar collects in the central region a mass of gas that is a significant
fraction of the dynamical mass, then the gaseous disk might become
dynamically unstable and form a gaseous bar that could drive gas further into
the nuclear region. A nuclear gaseous bar has been recently
discovered in the nearby Circinus galaxy (Maiolino et al. 1998b),
that hosts a Compton thick Seyfert 2
nucleus, in agreement with expectations. Ironically, this galaxy has been
classified as non-barred in the RC3 catalog; however the fact that it
is edge on and located in the Galactic plane (where crowding and
patchy extinction confuse the large scale morphology) might have
prevented the identification of a large scale stellar bar.
Alternatively, nested secondary {\it stellar} bars have also been observed
in several galaxies and also in Seyfert galaxies (Mulchaey \& Regan 1997),
and are thought to be more stable (Friedli \& Martinet 1993).
An encouraging result in this direction is
 that the only two Sy2s showing evidence for
an inner secondary stellar
bar and for which the N$_H$ has been estimated are actually
heavily obscured (see Tab. 1).
However the opposite is not true: 5 other
Compton thick Sy2s imaged in the near--IR by Mulchaey et al. (1997) do not
show evidence for inner bars, though the limited angular resolution might
have prevented their detection.

\begin{acknowledgements}
We are grateful to L. Hunt for providing
us with information on her data in advance of publication.
This work was partially supported by the Italian Space Agency (ASI)
through the grant ARS--98--116/22.
\end{acknowledgements}

\end{document}